\documentclass[a4paper]{PoS}
\usepackage{graphicx}
\usepackage{epstopdf}
\usepackage{subcaption}
\title{Cherenkov Telescope Array potential in the search for Galactic PeVatrons}

\ShortTitle{CTA PeVatrons}

\author{\speaker{E. O. Ang\"uner}$^{a}$, F. Cassol$^{a}$, H. Costantini$^{a}$, C. Trichard$^{b}$ and G. Verna$^{a}$ for the CTA Consortium\footnote{for collaboration list see PoS(ICRC2019)1177}.\\
\llap{$^a$} Aix Marseille Univ, CNRS/IN2P3, CPPM, Marseille, France\\
\llap{$^b$} LLR CNRS/IN2P3, F-91128 Palaiseau, France\\

E-mail: \email{oanguner@cppm.in2p3.fr} \\}

\abstract{One of the major scientific objectives of the future Cherenkov Telescope Array (CTA) Observatory is the search for PeVatrons. PeVatrons are cosmic-ray factories able to accelerate nuclei at least up to the knee feature seen in the spectrum of cosmic rays measured near the Earth. CTA will perform a survey of the full Galactic plane at TeV energies and beyond with unprecedented sensitivity. The determination of efficient criteria to identify PeVatron candidates during the survey is essential in order to trigger further dedicated observations. Here, we present results from a study based on simulations to determine these criteria. The outcome of the study is a PeVatron figure of merit, defined as a metric that provides relations between spectral parameters and spectral cutoff energy lower limits. In addition, simulations of the PeVatron candidate HESS\,J1641$-$463 and its parental particle spectrum are presented and discussed. Eventually, our work is applied to simulated population of Galactic PeVatrons, with the aim to determine the sensitivity of CTA.}

\FullConference{36th International Cosmic Ray Conference -ICRC2019-\\
		July 24th - August 1st, 2019\\
		Madison, WI, U.S.A.}

\begin{document}


\section{Introduction}

Cosmic rays (CRs) are primarily energetic nuclei, dominated by protons, and their spectrum observed at Earth shows a major spectral feature called the \emph{knee}, where the differential power-law spectrum steepens from $E^{-2.7}$ to $E^{-3.0}$ at an energy of a few Peta-electronvolt (1 PeV = 10$^{15}$\,eV). Note that the knee feature could be the result of different knee-like features seen at increasing CR energies for increasing atomic number. The recent results showed that the location of the knee in the observed proton and helium spectra is at 400--500\,TeV \cite{bartoli2015}. Cosmic rays up to the knee are believed to have originated in our Galaxy. In order to maintain the CR intensity at the observed level, the CR sources in our Galaxy must provide $\sim$10$^{41}$\,erg/s in the form of accelerated particles \cite{crs}. 

Galactic PeVatrons are the CR factories able to accelerate nuclei at least up to PeV energies. Supernova remnants (SNRs) are promising PeVatron candidates since they satisfy the energetic requirements if they can convert $\sim$10$\%$ of their kinetic energy into accelerated particles \cite{snr}, thus may accelerate particles up to PeV energies. In 2016, the H.E.S.S. Collaboration reported on the discovery of first Galactic PeVatron in the Galactic Center (GC) region \cite{gcpev} within a few pc around the super-massive black hole Sgr A*, based on a spectral investigation of the diffuse very-high-energy (VHE, E $>$ 0.1 TeV) $\gamma$-ray emission around the GC. Such a detection opens up new possible scenarios for the acceleration of PeV CRs in our Galaxy and indicates that not only SNRs but also other type of astrophysical sources can be PeVatrons in our Galaxy. A recent study suggests that star-burst regions with multiple powerful winds of young massive stars \cite{stecul} can also be promising PeVatron candidates. 

The PeVatron sources are expected to have hard power-law spectra (i.e. not much steeper than $E^{-2}$) that extend up to 50 TeV energies and beyond. Note that a firm detection of 100\,TeV intrinsic spectral cutoff will most likely make a source a PeVatron. In such case, at least 1 PeV CRs would be needed at the source region to explain observed VHE emission, assuming a conservative proton\,/\,$\gamma$-ray energy ratio of 10 \cite{kelner}. 

The fundamental question of the origin of the cosmic-ray sea that fills our Galaxy, especially where $\&$ how CRs are accelerated up to PeV energies and the distribution of PeVatron sources in our Galaxy, is still open.

\section{Cherenkov Telescope Array potential in PeVatron search} 

The Cherenkov Telescope Array is the next generation gamma-ray observatory consisting of an array of more than hundred imaging atmospheric Cherenkov telescopes (IACTs). CTA will be able to observe the entire sky. The array is located in two sites, one at the Southern and another at the Northern hemisphere. The Southern array will include three different telescope types/sizes to maximize the covered energy range and will be able to provide measurements of VHE $\gamma$-rays up to $\sim$300\,TeV. The expected sensitivity of Southern array above tens of TeV is several orders of magnitude better than current IACT experiments \cite{ksp}. This indeed motivates for deep PeVatron searches with CTA, since the Southern array will be the right instrument to find the origin of CRs.

One of the major scientific objectives of the CTA Observatory is the search for cosmic PeVatrons. It is, in fact, one of the science cases covered by key science projects of the CTA consortium. The goal is to address some of the fundamental questions of high energy CR acceleration and provide information on the distribution of PeVatrons in our Galaxy \cite{ksp}.

The Southern array of the CTA will perform a complete Galactic Plane Survey (GPS), providing an 2--4\,mCrab\footnote{Throughout this study, 1 Crab flux at 1\,TeV is taken as $\Phi_{Crab}$ (1 TeV) = 3.8 $\times$ 10$^{-11}$ cm$^{-2}$ s$^{-1}$ TeV$^{-1}$ \cite{crabpaper}. The GPS sensitivity of 2--4 mCrab is for an energy threshold of 0.125\,TeV.} survey sensitivity and an average 10--15 h exposure depending on Galactic coordinates \cite{gps}. The CTA GPS will provide an unprecedented dataset for the search of Galactic PeVatrons. The PeVatron search program proposes 50 h deeper observations of the best five PeVatron candidates selected from the CTA GPS. Thus, the determination of efficient criteria to identify PeVatron candidates during the survey is essential in order to trigger further dedicated observations. In this study, we present results from a study based on simulations to determine these criteria, which can lead to the selection of sources having the highest intrinsic cutoff in their spectra. 

\section{CTA simulations and data analysis}

Throughout this study, we simulated point-like PeVatron sources together with the CR background events. The PeVatron source spectra are parameterized both by power-law (PL) and exponential cutoff power-law (ECPL) models. The simulations\footnote{Both the simulations and analysis of the simulated data are performed with CTOOLS\cite{ctools} by using prod3b-v1\cite{irf} instrument response functions at zenith angle of 20$^\circ$ and offset angle of 0.7$^\circ$.} are performed for a set of range of PL parameters, spectral index of $\Gamma$ = $-$1.7, $-$2.0 and $-$2.3 and flux normalization at 1 TeV of $\Phi_{0}$\,=\,4,\,8,\,16, 24, 32, 40 and 48 mCrab, assuming conservative 10 h of GPS observation time. For the simulation of ECPL sources, intrinsic spectral cutoff energies (E$_{c,\gamma}$) of 50\,TeV, 100\,TeV, and 200\,TeV are assumed in addition to the given set of PL parameters. The energy range of the simulations is between 0.1\,TeV and 160\,TeV. The simulations are repeated 1000 times for each set of $\Gamma$, $\Phi_{0}$ and E$_{c,\gamma}$. The phase space covering the simulations will be referred to as PeVatron phase space. In addition, we simulated 50 h of CTA observations of the PeVatron candidate source HESS\,J1641$-$463 \cite{igor} for the investigation of its parental proton spectrum characteristics.  

Analysis of the simulated sources is performed using the reflected background estimation method \cite{refbg}. The simulated data are fitted both with PL and ECPL models maximizing the ON-OFF Poisson likelihood function, referred to as the $W$ statistic in XSPEC \cite{xspec}. The test statistics (TS = $W_{ECPL}$ - $W_{PL}$) is used to determine the spectral model that describes the simulated data at best\footnote{An ECPL model is preferred over a PL model if TS $\geq$ 9.0, which corresponds to a 3$\sigma$ confidence level. }. The 95$\%$ confidence level (C.L.) lower limits on the cutoff energy (E$_{c,\gamma}^{95\%}$) are derived for the sources in which the cutoff can not be firmly detected (TS $<$ 9.0). We use two different approaches for the derivation of the E$_{c,\gamma}^{95\%}$. In the first approach, the 95\% C.L. lower limits are obtained from the fitted cutoff distribution of 1000 simulations, taking the 5 percentile of the distribution as the E$_{c,\gamma}^{95\%}$. In the second one, the profile likelihood method described in \cite{cyril} is used.

\section{PeVatron candidate source HESS\,J1641$-$463}

One of the most promising PeVatron candidate is HESS\,J1641$-$463 \cite{igor}. The source is point-like and exhibits a relatively hard spectrum, extending to a few tens of TeV without showing any clear sign of a cutoff. We simulated 50 h of CTA observations of this source 1000 times, assuming 50\,TeV, 100\,TeV and 200\,TeV intrinsic cutoff in its spectrum, while keeping $\Gamma$ = -2.07 and $\Phi_{0}$\,(1\,TeV)\,=\,3.91 $\times$ 10$^{-13}$ cm$^{-2}$ s$^{-1}$ TeV$^{-1}$ ($\sim$18 mCrab above 1 TeV) fixed at the published values. 

\begin{figure}[!ht]
\centering
\includegraphics[width=1.0\textwidth]{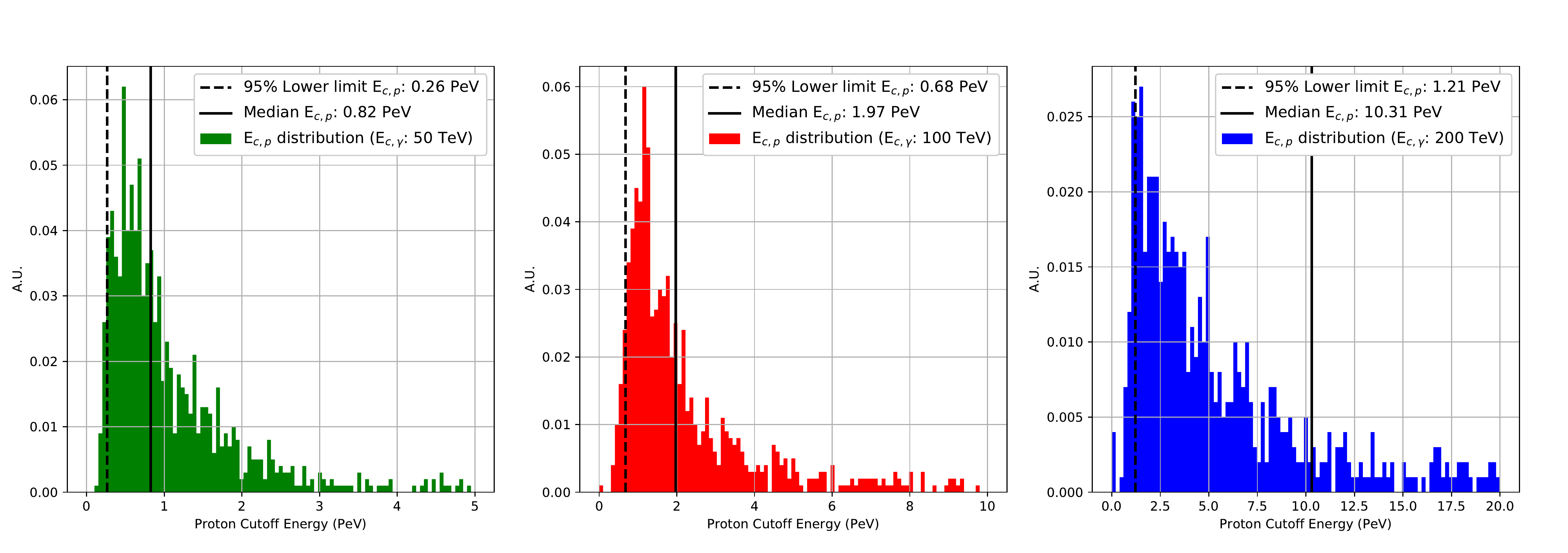}
\caption{The distributions of fitted parental proton cutoff energies assuming intrinsic E$_{c,\gamma}$ = 50\,TeV (left), E$_{c,\gamma}$ = 100\,TeV (middle) and E$_{c,\gamma}$ = 200\,TeV (right) for the PeVatron candidate source HESS\,J1641$-$463. The dashed and solid lines show the 95\% C.L. lower limits on the E$_{c,p}$ and median of the distributions, respectively.}
\label{proton}
\end{figure}

Flux points derived from each simulation are given to the Naima package \cite{naima} as input to obtain parameters of the parental proton spectra described by the ECPL spectral model. The interstellar medium properties published in \cite{igor}, n$_{gas}$ of 100 cm$^{-3}$ and distance of 11 kpc, are used for the hadronic modeling of the source. The parental proton spectra cutoff energy (E$_{c,p}$) probability distributions obtained by assuming intrinsic E$_{c,\gamma}$ of 50\,TeV, 100\,TeV and 200\,TeV are shown in Fig.\,\ref{proton}.

Note that for the case of intrinsic E$_{c,\gamma}$ of 50\,TeV, the median ($\sim$0.8 PeV) and the 95\% C.L. lower limit on the E$_{c,p}$ ($\sim$0.3 PeV) values obtained from the parental proton E$_{c,p}$ distribution (Fig. \ref{proton} left) are comparable to the reported knee value of 400--500\,TeV \cite{bartoli2015} in the observed proton and helium spectra. Thus, these preliminary results suggest that even in the case of 50\,TeV intrinsic cutoff, together with a spatial coincidence of dense gas regions, HESS\,J1641$-$463 (or similar hard spectrum sources) can contribute to the knee in the observed proton and helium spectra. Moreover, for the cases of intrinsic E$_{c,\gamma}$ = 100\,TeV and E$_{c,\gamma}$ = 200\,TeV (Fig. \ref{proton} middle and right), the median ($\sim$2.0\,PeV and $\sim$10.0\,PeV) and the 95\% C.L. lower limit on the E$_{c,p}$ ($\sim$0.7\,PeV and $\sim$1.2\,PeV) values are comparable to the knee structure seen at an energy of a few PeV. Thus, the detection of such high intrinsic cutoff energies can lead to the firm detection of Galactic PeVatrons. For this reason, we use the set of E$_{c,\gamma}$ = 50 TeV, 100 TeV and 200 TeV for the creation of the PeVatron phase space. Recall that the reported 95\% C.L. lower limit on the E$_{c,p}$ for the Galactic Center PeVatron is 0.4\,PeV \cite{gcpev}. 

\section{Spectral cutoff detection maps}  

\begin{figure}[!h]
\begin{subfigure}{.5\linewidth}
\centering
\includegraphics[width=1.0\textwidth]{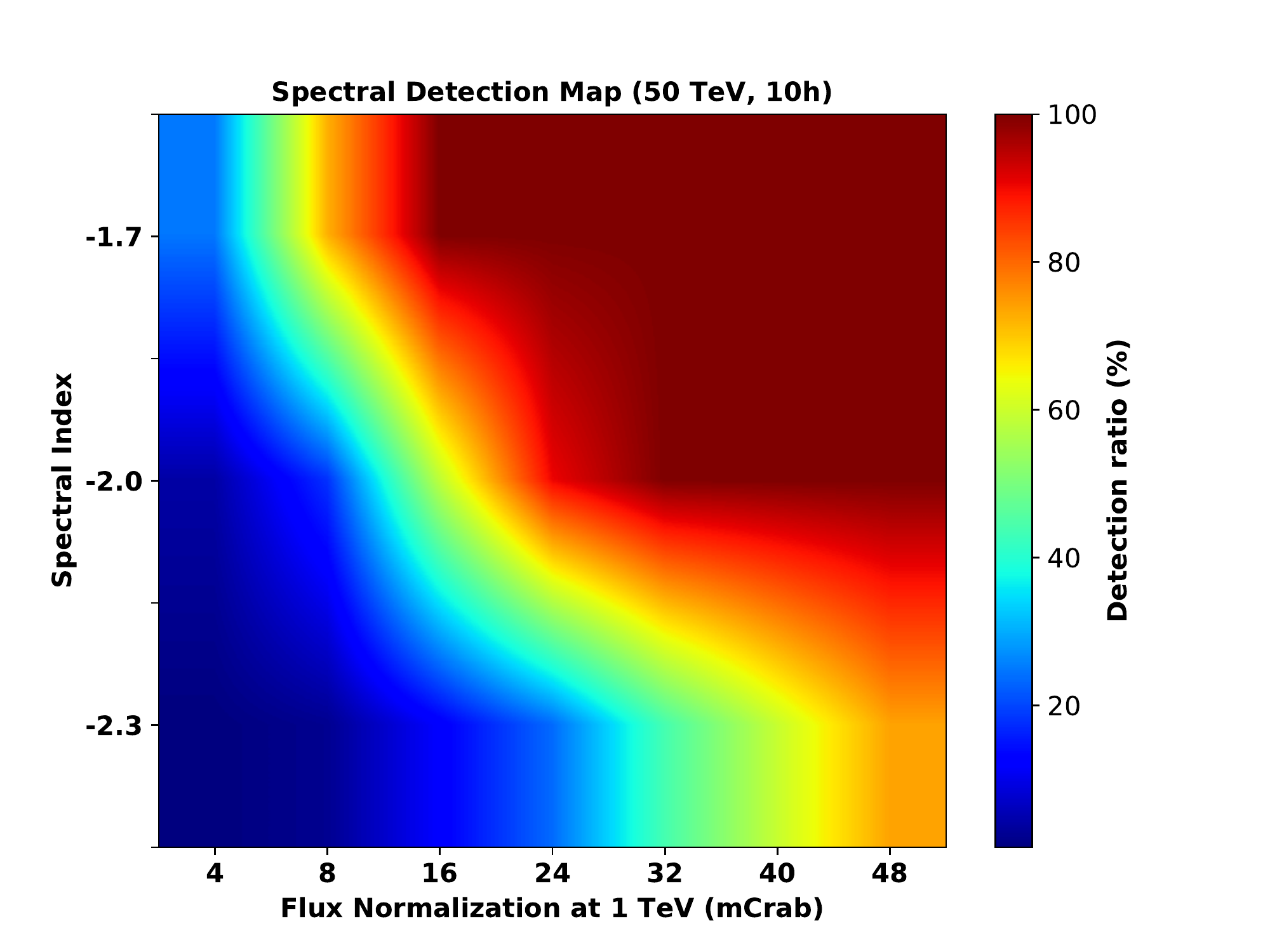}
\caption{}
\label{fig:sub1}
\end{subfigure}%
\begin{subfigure}{.5\linewidth}
\centering
\includegraphics[width=1.0\textwidth]{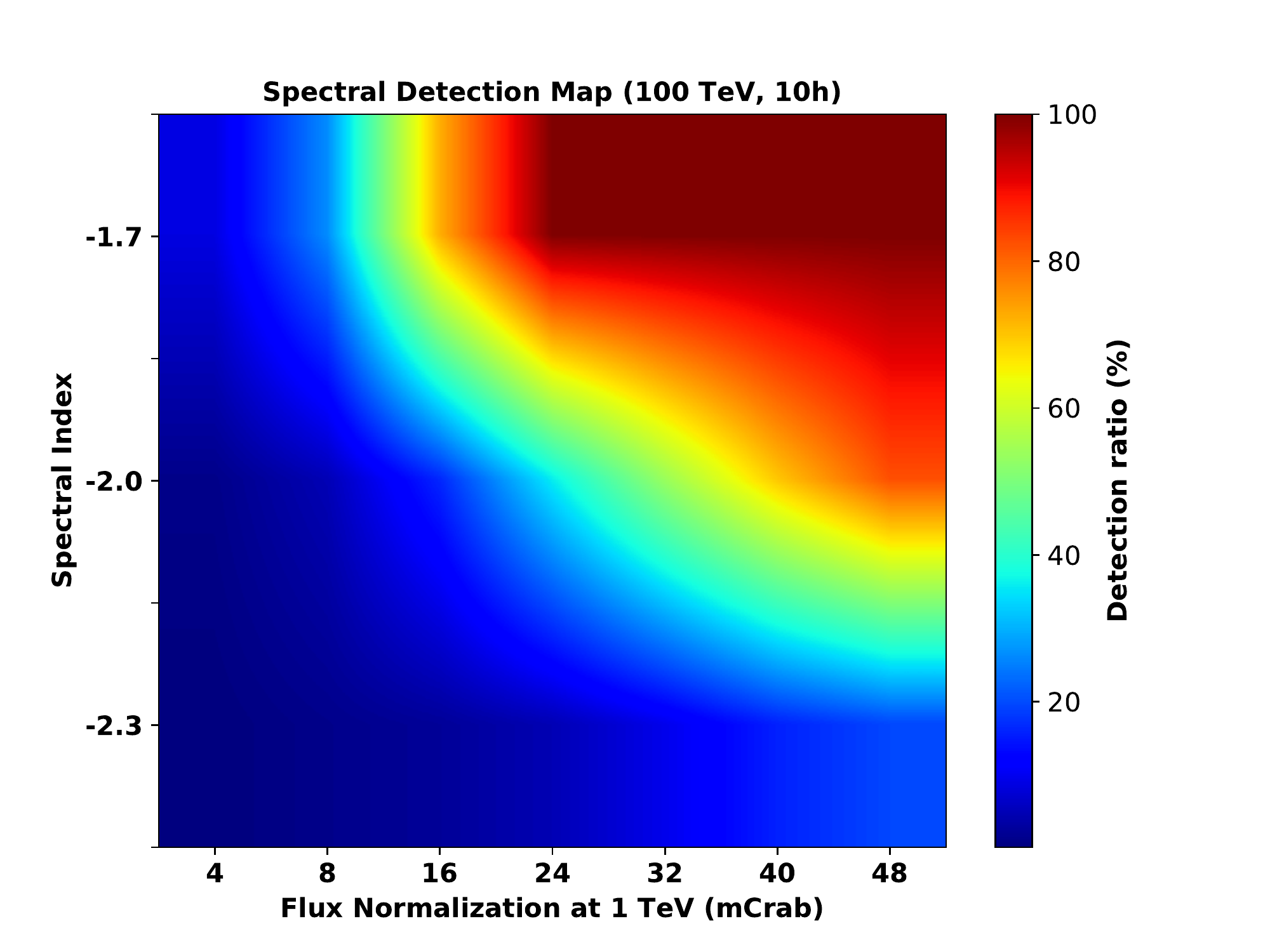}
\caption{}
\label{fig:sub2}
\end{subfigure}\\[1ex]
\begin{subfigure}{\linewidth}
\centering
\includegraphics[width=0.5\textwidth]{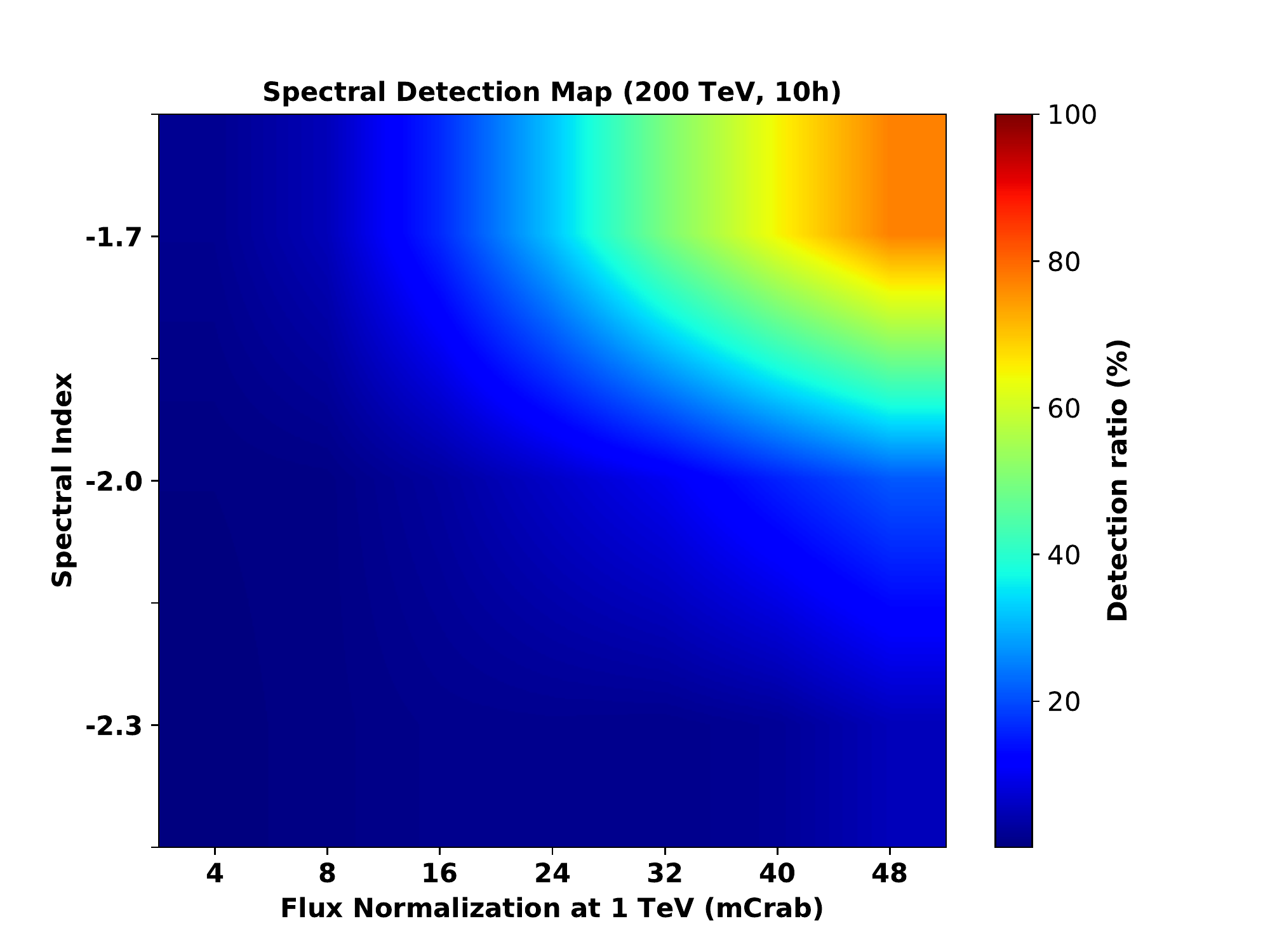}
\caption{}
\label{fig:sub3}
\end{subfigure}
\caption{The spectral cutoff detection maps of E$_{c,\gamma}$\,=\,50\,TeV (a), E$_{c,\gamma}$\,=\,100\,TeV (b) and E$_{c,\gamma}$\,=\,200\,TeV (c). The color bar shows the percentage of simulations for which the cutoff can be firmly (TS $>$ 9.0) detected. The detection percentage shown in each bin of the map is obtained considering 1000 simulated spectra for each specific spectral model ($\Gamma$, $\Phi_{0}$, E$_{c,\gamma}$). We use bilinear interpolation for smoothing the maps.}
\label{detmap}
\end{figure}

In this section we estimate the probability to detect a spectral cutoff of sources having E$_{c,\gamma}$ of 50\,TeV, 100\,TeV and 200\,TeV during the CTA GPS after a conservative 10 h observation time. The spectral cutoff detection maps are shown in Fig. \ref{detmap}.

Note that in general, the detection probability of higher intrinsic E$_{c,\gamma}$ increases with source brightness and/or as source spectrum gets harder. This is basically due to the fact that these effects provide increased statistics at the high energy part of the source spectrum, resulting in a better description of the spectral shape. 

Figure \ref{detmap} (a) and (b) show that the cutoff of point sources with intrinsic E$_{c,\gamma}$ of 50 TeV and 100 TeV can be detected in the case of hard  spectral index at the CTA GPS sensitivity level. The detection of intrinsic E$_{c,\gamma}$ = 200 TeV during the survey may be possible for very hard and bright sources. 


\section{PeVatron metric}

\begin{figure*}[!ht]
\begin{subfigure}{.5\linewidth}
\centering
\includegraphics[width=1.1\textwidth]{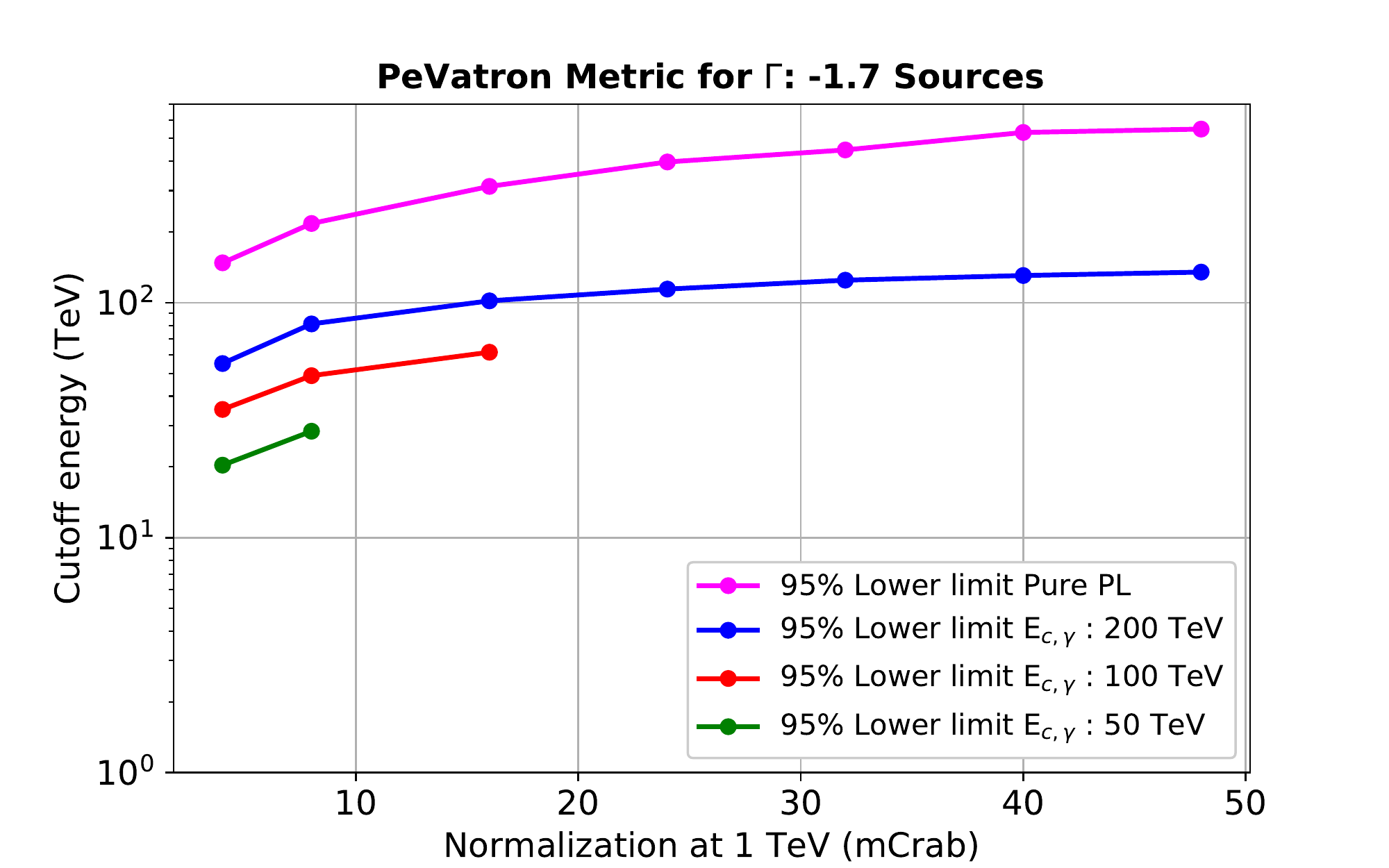}
\caption{}
\label{fig:sub1}
\end{subfigure}%
\begin{subfigure}{.5\linewidth}
\centering
\includegraphics[width=1.1\textwidth]{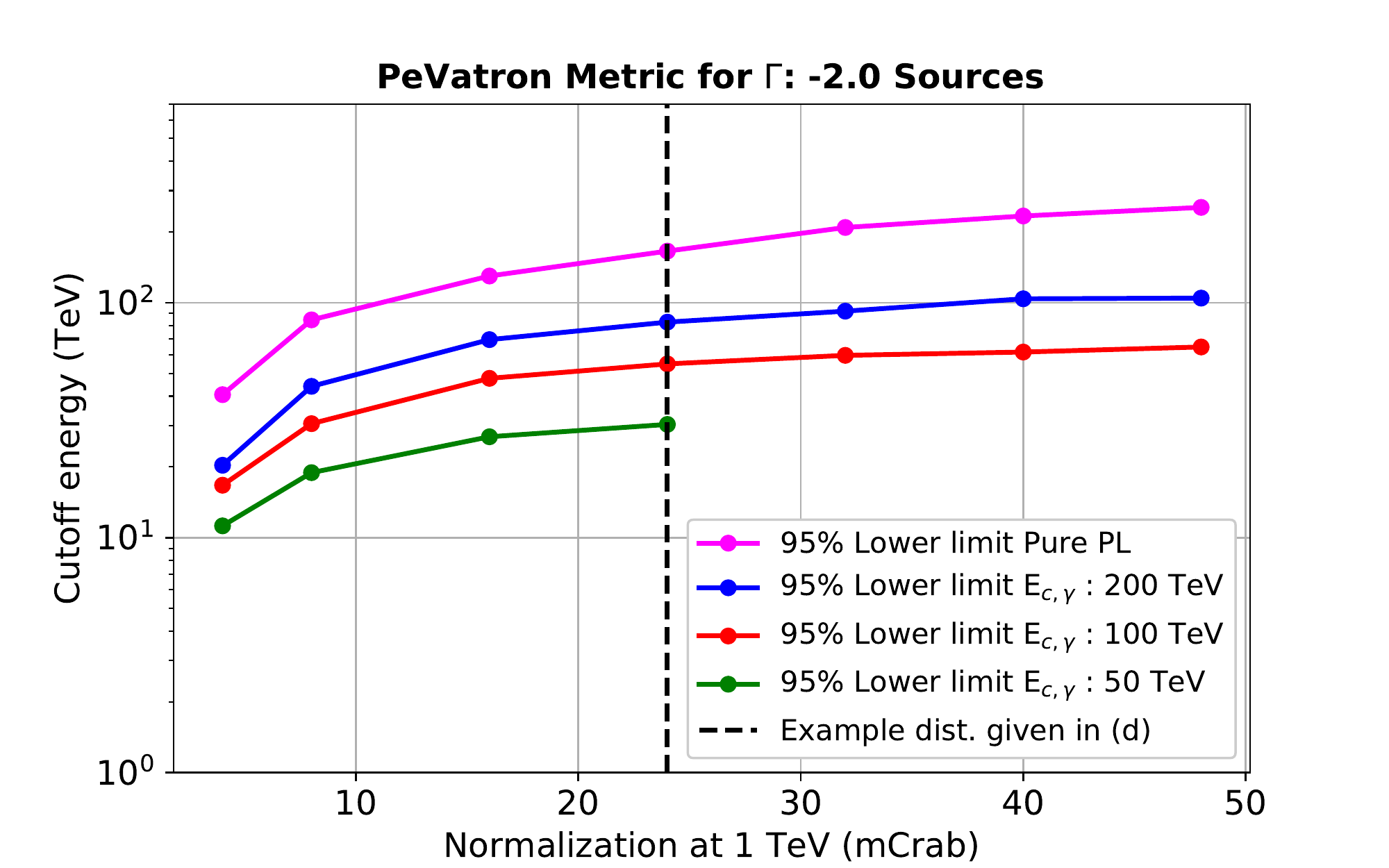}
\caption{}
\label{fig:sub2}
\end{subfigure}
\begin{subfigure}{.5\linewidth}
\centering
\includegraphics[width=1.1\textwidth]{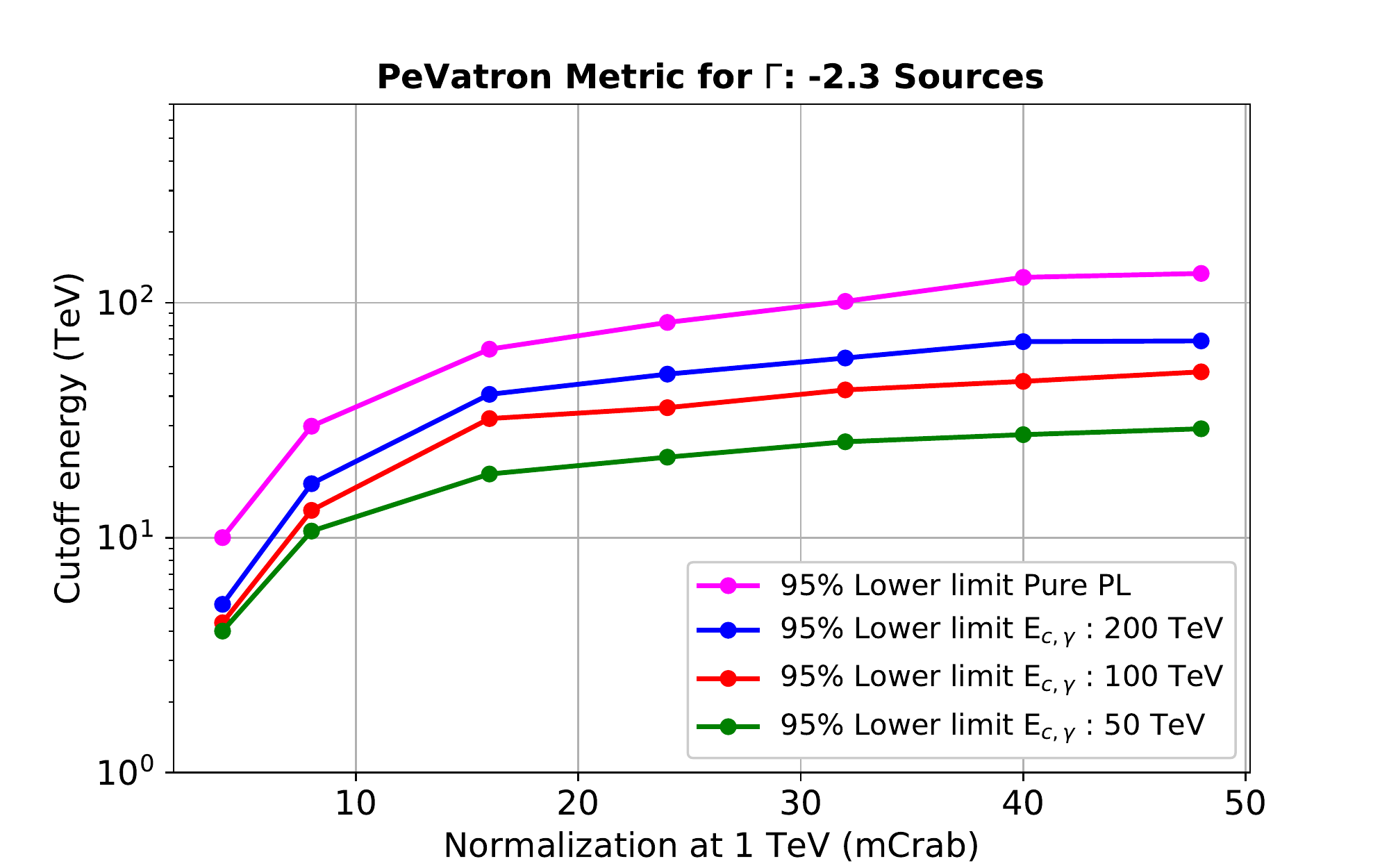}
\caption{}
\label{fig:sub3}
\end{subfigure}
\begin{subfigure}{.53\linewidth}
\centering
\includegraphics[width=1.0\textwidth]{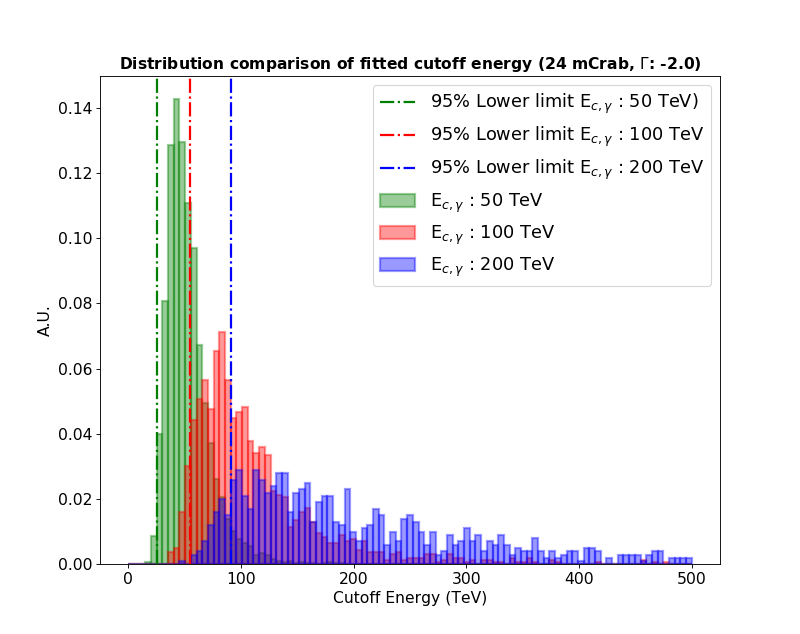}
\caption{}
\label{fig:sub4}
\end{subfigure}
\caption{The PeVatron metric for sources with spectral index of $-$1.7\,(a), $-$2.0\,(b) and $-$2.3\,(c). The E$_{c,\gamma}^{95\%}$ for pure PL sources are shown with the magenta solid lines. The E$_{c,\gamma}^{95\%}$ derived for sources with intrinsic E$_{c,\gamma}$ = 50 TeV, E$_{c,\gamma}$ = 100 TeV and E$_{c,\gamma}$ = 200 TeV are shown with the green, red and blue solid lines, respectively. An example of fitted cutoff probability distributions for a source with $\Gamma$: $-$2.0 and $\Phi_{0}$\,=\,24\,mCrab is shown in (d).}
\label{pevmet}
\end{figure*}

In this section, we introduce a concept called PeVatron metric, which is a figure of merit for PeVatron candidate sources. The metric can provide relations between spectral parameters ($\Gamma$,\,$\Phi_{0}$,\,E$_{c,\gamma}$) and derived E$_{c,\gamma}^{95\%}$ for the sources located within the PeVatron phase space. The metric is produced for each set of spectral parameters, based on 1000 simulations for which the intrinsic cutoff cannot be detected. The E$_{c,\gamma}^{95\%}$, shown by solid lines in Fig. \ref{pevmet}, are obtained by taking the 5 percentile of the fitted cutoff probability distributions as shown in Fig. \ref{pevmet} (d).

Note that the metric lines stop at the flux level that corresponds to 100\% detection level (e.g. \,$\Gamma$:\,$-$1.7, E$_{c,\gamma}$:\,50\,TeV and $\Phi_{0}$ $>$ 8\,mCrab) in some cases. This is in agreement with the spectral cutoff detection maps given in Fig. \ref{detmap}. It can be seen that the E$_{c,\gamma}^{95\%}$ increases as the source gets brighter and/or the source spectrum gets harder. Moreover, the E$_{c,\gamma}^{95\%}$ for sources having identical $\Gamma$ and $\Phi_{0}$ increase as the intrinsic E$_{c,\gamma}$ gets higher (see Fig. \ref{pevmet} (d)). This fact can actually be used for predicting the intrinsic E$_{c,\gamma}$ of a source of interest. 

\section{Large scale simulations}

The prediction of intrinsic E$_{c,\gamma}$ from the data collected from CTA GPS is essential for ranking the sources and selecting the best PeVatron candidates. For this purpose, we performed a large scale simulations with 1000 sources to simulate the PeVatron phase space. The sources are simulated by following ECPL models, with randomly generated parameters within the phase space, assuming 10\,h of observation time. Note that the intrinsic E$_{c,\gamma}$ cannot be detected for any of these sources. We calculated the total number of excess events above 50 TeV in addition to the E$_{c,\gamma}^{95\%}$ for each source. The E$_{c,\gamma}^{95\%}$ are derived by creating 1000 fake spectra for each source and taking the 5 percentile of the fitted cutoff distribution. The fake spectra are created from the observed N$_{On}$ \& N$_{Off}$ events from each source spectrum, by scrambling N$_{On}$ \& N$_{Off}$ events under the assumption of Poisson distribution.

\begin{figure*}[!ht]
\begin{subfigure}{.5\linewidth}
\centering
\includegraphics[width=1.1\textwidth]{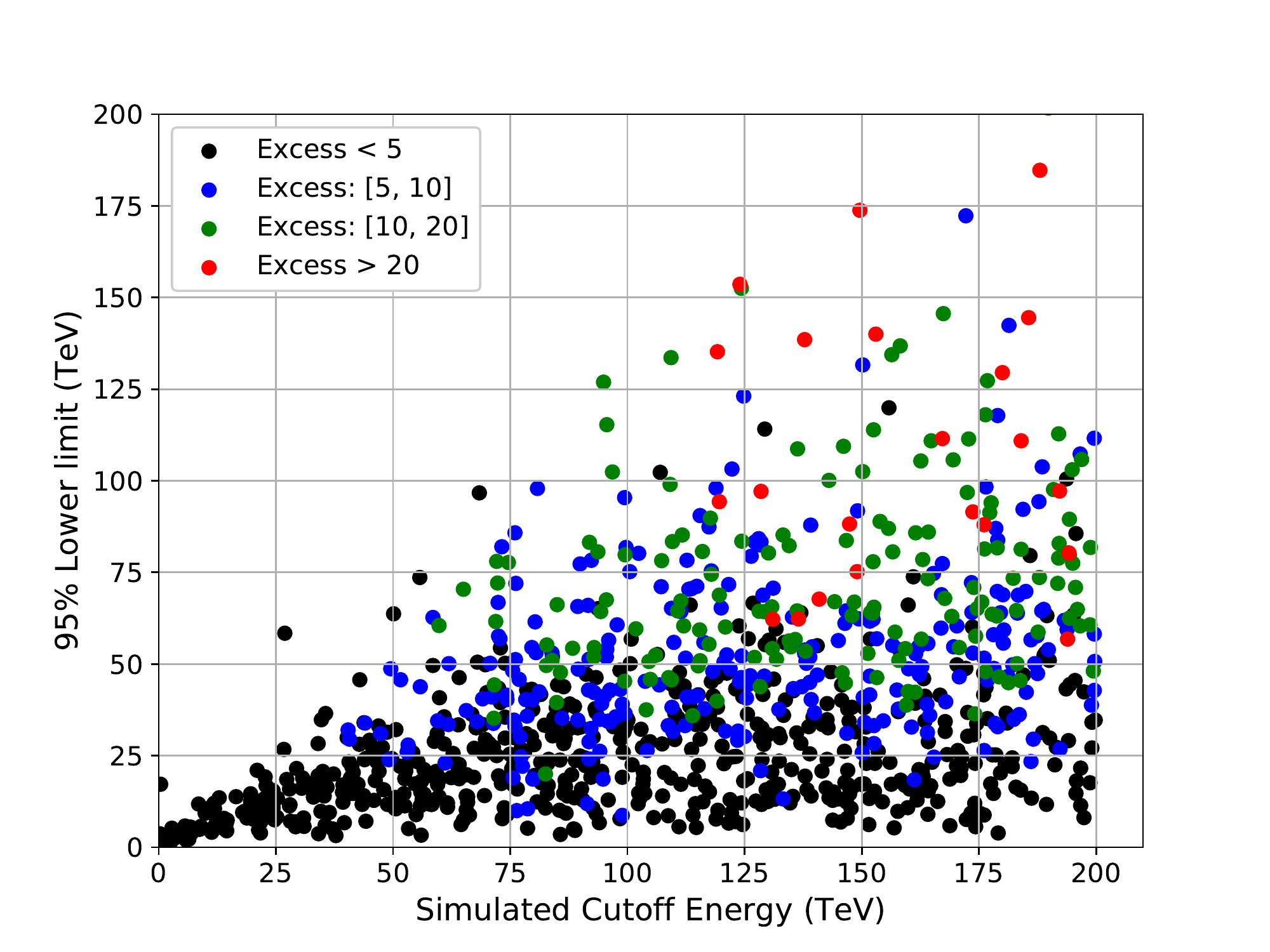}
\caption{}
\label{fig:sub1}
\end{subfigure}%
\begin{subfigure}{.5\linewidth}
\centering
\includegraphics[width=1.1\textwidth]{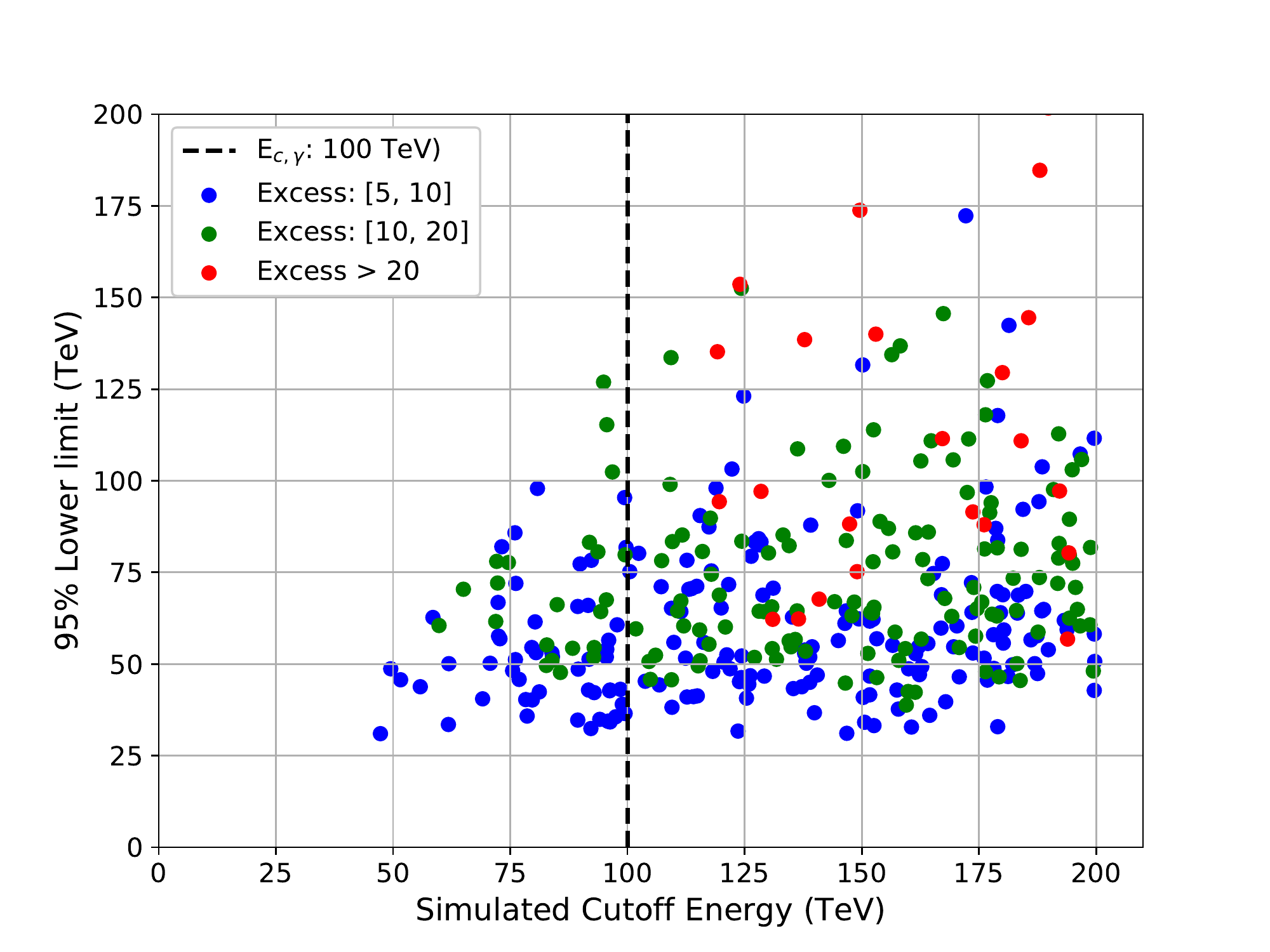}
\caption{}
\label{fig:sub2}
\end{subfigure}
\caption{The distribution of 1000 test sources, categorized by using the total number of excess events above 50 TeV, before (a) and after (b) the application of the metric.}
\label{lsrt}
\end{figure*}

Figure \ref{lsrt} (a) shows the distribution of the sources, categorized by using the total number of excess events above 50 TeV. One can see that there is a strong correlation between the derived E$_{c,\gamma}^{95\%}$ and the total number of high energy excess events. Thus, both parameters are useful and look promising for defining the final selection criteria. From the plot, one can see that no prediction on the intrinsic E$_{c,\gamma}$ can be obtained in the case of low E$_{c,\gamma}^{95\%}$ and/or low excess events at high energies. Such sources are not promising candidates for dedicated observations and can be ruled out.

It is possible to use the PeVatron metric for the prediction of intrinsic E$_{c,\gamma}$. One can obtain the expected E$_{c,\gamma}^{95\%}$ for each source by interpolating between the metric lines. The fitted $\Gamma$ and $\Phi_{0}$ parameters are given to the metric as input. The output parameters are the expected E$_{c,\gamma}^{95\%}$ values for intrinsic E$_{c,\gamma}$ of 50 TeV, 100 TeV and 200 TeV. First, we applied a metric selection criterion by restricting the derived E$_{c,\gamma}^{95\%}$ for a source that should be higher than the expected E$_{c,\gamma}^{95\%}$ for the case of E$_{c,\gamma}$\,=\,100 TeV. With this criterion, we choose only the sources with expected intrinsic E$_{c,\gamma}$\,$\geq$\,100\,TeV. Second, we excluded the sources having low excess events above 50 TeV since there is not enough statistics to make a prediction on their intrinsic E$_{c,\gamma}$. The outcome distribution of selected sources is shown in Figure \ref{lsrt} (b). Note that most of the unfavorable sources having low intrinsic E$_{c,\gamma}$ have been ruled out while keeping most of the promising sources. Performing the simulations for each source by using identical $\Gamma$ and $\Phi_{0}$ obtained from the source data (instead of interpolation from the metric) should provide more accurate predictions. Dedicated studies on the prediction of intrinsic E$_{c,\gamma}$ of each individual source are going on.  

\section{Conclusions}

Our simulation studies suggest that intrinsic spectral cutoffs of 50 TeV and 100 TeV can be detected especially for hard sources during the CTA GPS even for a short observing time of 10 h. We show that the 95\% C.L. lower limit on the cutoff energy increases with source brightness and/or as source spectrum gets harder. In addition, it is shown that the E$_{c,\gamma}^{95\%}$ increases with intrinsic E$_{c,\gamma}$ of the source of interest. Preliminary investigation show that indications on the intrinsic E$_{c,\gamma}$ can be estimated using the E$_{c,\gamma}^{95\%}$ and the excess events at high energies. Further studies are needed to derive robust selection criteria for PeVatron candidates. 

\acknowledgments

This work was conducted in the context of the CTA Consortium. We gratefully acknowledge financial support from the agencies and organizations listed here:

\texttt{http://www.cta-observatory.org/consortium\_acknowledgments}, and especially the OCEVU Labex (ANR-11-LABX-0060) and the Excellence Initiative of Aix-Marseille University - A*MIDEX, both part of the French ``Investissements d'Avenir'' programme.

\end{document}